# Monte-Carlo Multiscale simulation study of argon adsorption / desorption hysteresis in mesoporous heterogeneous tubular pores like MCM-41 or oxidized porous silicon

*Joël Puibasset*

Centre de Recherche sur la Matière Divisée, CNRS-Université d'Orléans, 1b rue de la Férollerie, 45071 Orléans, Cedex 02, France.

puibasset@cnrs-orleans.fr

**Monte-Carlo Multiscale simulation**.

Corresponding author. E-mail: puibasset@cnrs-orleans.fr

ABSTRACT

In a recent paper [J. Chem. Phys. **127**, 154701 (2007)], a multiscale approach was introduced which allowed the calculation of adsorption/desorption hysteresis for fluid confined in a single mesoporous, heterogeneous tubular pore. The main interest in using such an approach is that it allows to reconcile a molecular simulation approach generally limited to the nanometer scale (atomistic description of the confined fluid and pore roughness) with the much larger scale (micrometer) relevant to understand the complexity of adsorption/desorption hysteresis (the numerous metastable states in the hysteresis loop are a consequence of the large scale disorder in the porous material). In this paper, this multiscale approach is used to study adsorption phenomena in mesoporous models made of a collection of disordered, non-interconnected tubular pores, as MCM-41 or porous silicon. A double-distribution is introduced, one to characterize the disorder in a given pore, and the other one to characterize the disorder between the pores. We consider two distribution shapes: Gaussian and uniform truncated, and the two cases of pores open at one or at both ends. These models are expected to cover a wide variety of real materials made of independent pores, as MCM-41 and oxidized porous silicon. A large variety of hysteresis shapes is obtained, ranging from almost parallel adsorption/desorption branches typical of MCM-41 adsorption, to triangular hysteresis typical of porous silicon. The structure of the metastable states inside the hysteresis (scanning adsorption/desorption curves) is also examined. The results are expected to be useful to experimentalists who want to infer pore structure and level of disorder from experimental adsorption/desorption experiments.

# INTRODUCTION

An essential feature of porous material is its structure, which is obviously easier to take into account to determine material properties when geometry is simple. This explains the importance of inorganic mesoporous materials of well defined geometry. In particular MCM-41 molecular sieves[1,2] and porous silicon,[3] which exhibit cylinderlike pore shape, and negligible interconnections between pores. These exceptional geometric properties and the fact that they are easy to produce explain they are commonly used for phase separation, catalysis, chemical sensors, and nanotechnologies. But they are also of greatest importance for fundamental studies on adsorption phenomena.[4-9] In this paper we focus on such porous materials which are believed to be made of a collection of independent (non-interconnected) pores. The case of perfectly cylindrical pores is quite well understood from a theoretical point of view.[10] However, there is still some disagreement between experimental observations of adsorption/desorption hysteresis and their theoretical prediction, showing that there is still room for improvement of pore structure description.

Despite exceptional geometric qualities, the above mentioned materials most probably contain some disorder, as suggested by the experimental adsorption/desorption hysteresis. For instance, a non-negligible pore size distribution exists (in particular in porous silicon), which can be confirmed by microscopy. Furthermore, the MCM-41 or porous silicon inner surfaces are most probably rough and irregular.[8,11-16] The consequence is a variation of pore size and wall thickness in the sample, which is a source of disorder.[17] The term disorder refers to any spatial variation in the porous material characteristics which has consequences on the thermodynamic properties of the fluid. It was shown elsewhere[18-20] that a purely geometric deformation of a pore (keeping the fluid/wall interaction constant) has negligible consequences on the thermodynamic properties of the adsorbed fluid. On the other hand, any modulation of fluid/wall interaction has strong consequences.[21-28] Of course, in a real porous material both geometric and fluid/wall effects cannot be disentangled, since pore size variations have direct consequences on fluid/wall interaction intensity. Important geometric effects are also expected when the structural disorder of the surface is transferred to the adsorbed fluid phase. Such effects are neglected in this study and could constitute points for future improvement. In the present work, it will be considered that the most important sources of disorder are those which affect directly the fluid/wall interaction. The porous substrates considered in this study exhibit three important sources of disorder, associated to three length scales. The shortest wavelength variations in fluid/wall interactions are due to the atomistic roughness of the walls. This affects essentially the first stage of fluid adsorption (from low chemical potential up to the monolayer) by creating favorable sites for adsorption. This is naturally taken into account in our molecular simulation approach, provided a reasonable atomistic pore model is used. Longer wavelength variations also exist. As previously mentioned, for a given pore, one has possible chemical or wall thickness variations along the pore axis which induces modulation of the fluid/wall interactions: the pore is heterogeneous. The typical scale for such heterogeneities is the nanometer. It corresponds to the size of the elemental domains of the material, as introduced by Everett.[29] This is probably the most important source of disorder to explain the hysteresis, as shown by the work of Kierlik and coworkers[30-32] in a mean-field density functional theory of a disordered lattice gas model. It is also the most complex to be modeled because these domains generally cannot be considered as independent. In most cases, one has to take into account the whole chain of domains of a given pore in order to calculate accurately its adsorption properties. This is achieved by the multiscale approach used in this work. It is emphasized that, in this work, the term "pore" should not be confused with "domain": the porous sample is seen as a collection of pores which are themselves made of a chain of domains. The last source of disorder is due to possible variations of global pore characteristics over the whole sample. For instance, inspection of porous silicon surface by microscopy shows a distribution of pore size which is wider than the variations of the size of any given pore along its axis.[33] This source of disorder can easily be taken into account in calculations since the pores are independent in MCM-41 and porous silicon.[7] This requires that the wall thickness is large enough to avoid strong interaction between molecules through the walls. It also requires that the pore structure is rigid enough not to be distorted during adsorption, because this would have non-negligible effects on fluid adsorption in the neighboring pores. These hypotheses are assumed.

The aim of this paper is to calculate the adsorption/desorption hysteresis as well as the first ascending and descending scanning curves for a simple fluid (e.g. argon) adsorbed in mesoporous silica, taking into account the three sources of disorder. Nanometer-scale and large-scale heterogeneities are modeled in a realistic but simplified way so as to be characterized by a single parameter. Two distributions are then introduced, one to characterize the variations between pores, the other one to model the domain heterogeneity in each pore. By varying the variance of these distributions, one may model either a collection of identical homogeneous pores, or a distribution of various homogeneous pores, or a collection of statistically identical but heterogeneous pores, or a large distribution of heterogeneous pores. The impact on hysteresis and scanning curves is studied and compared to experiments in order to infer conclusions on MCM-41 or porous silicon structure. The issue of pores open at one or at both ends is also addressed.

The paper is organized as follows: In section 2, the pore model is first introduced, and the two distributions characterizing the pore and domain heterogeneities are defined. The issue of system size effect is then discussed in order to define a reasonable number of pores and domains to avoid finite size artifacts. In section 3 the numerical details and algorithm are introduced, where it is shown how a multiscale approach may be successful to take into account the interconnections between the domains for each pore. The simulations results are given in section 4, followed by discussion and conclusion.

**PORE MODEL**

As explained in the introduction, a model for mesoporous silica as MCM-41 or oxidized porous silicon should describe the material as an arrangement of parallel and non-interconnected tubular pores with a high aspect ratio (few nanometers in diameter and few microns of length). Besides the atomistic disorder which is naturally taken into account in our molecular approach, some disorder should also be introduced, at two levels: differences between pores (for instance their average diameter may differ), and physico-chemical variations at nanometer scale along the axis of each pore. Note that these variations most probably rise during the fabrication of the material, and could in principle be obtained by numerical reconstructions.[17] With present computer capabilities, it is however impossible to reach the required large scales. A simplification is thus introduced, by considering the fact that the most relevant physico-chemical heterogeneities for fluid adsorption are those which induce a modulation of the fluid/wall interaction intensity (see introduction). As a consequence, the heterogeneities are characterized by a single parameter which varies from pore to pore and along any given pore. A more precise definition for this parameter is given below.

The silica pore diameter is taken to be 3 nm. This is in the range of the smallest MCM-41, but smaller than typical porous silicon pores (10nm or more), and results from a compromise due to computer limitations. The pore length is few micrometer, representative of mesoporous materials like MCM-41 or oxidized porous silicon In order to perform accurate calculations, the system is described at the molecular level, and the mesopores are drilled in an initial substrate made of pure silica of density close to the expected one (mesoporous silica walls are expected to be made of amorphous silica). The fluid/wall interaction intensity is then well reproduced since the density of interacting sites is correct. The fluid/fluid and fluid/substrate interactions are modeled by the (6-12) Lennard-Jones intermolecular potential cut and shifted at 2.5 $\sigma$, which is accurate enough to catch the main features of simple fluid adsorption. The parameters for argon are $\sigma_{Ar-Ar}$ = 0.3405 nm and $\varepsilon_{Ar-Ar}/k$ = 120K where $k$ is Boltzmann constant, and the parameters for argon/silica interactions are $\varepsilon^0_{Ar-O}/k$ = 100 K and $\sigma_{Ar-O}$ = 0.333 nm (fluid/wall interactions are mainly with O species). For further details concerning the choice for these typical values, see Ref [34]. As previously explained, the physico-chemical heterogeneities in the porous substrate are approximated by a modulation over the porous substrate of the fluid/wall interaction intensity $\varepsilon_{Ar-O}$ around its typical value $\varepsilon^0_{Ar-O}$, within ±30 percent. The exact value of $\varepsilon_{Ar-O}/\varepsilon^0_{Ar-O}$ over sample is of course sample-dependent and unknown. However, it is expected to have a typical correlation length (corresponding to the definition of a domain) and a random distribution at larger scale. It is expected that the domain size is few nanometers, and we have arbitrarily chosen the value 7.36 nm for the simulations. It is emphasized that choosing 5 nm or 15 nm would not change the results obtained in this work. A typical pore then contains several hundreds of domains. The heterogeneity parameter $h = \varepsilon_{Ar-O}/\varepsilon^0_{Ar-O}$ is supposed to be constant within each domain, and to take uncorrelated values between two adjacent domains. Its value within a given pore is supposed to follow a Gaussian distribution around its average value $h_{pore}$ in the given pore and variance $\sigma^2_{domain}$. The average pore value $h_{pore}$ is also supposed to follow a Gaussian distribution around 1.0 ($\varepsilon_{Ar-O} = \varepsilon^0_{Ar-O}$) and variance $\sigma^2_{pore}$. See Fig. 1 for a schematic representation. Since $h$ is restricted to the interval [0.7;1.3] the distributions are actually truncated, and normalized accordingly. The Gaussian distribution being strictly positive, all heterogeneity values between 0.7 and 1.3 may be found in all pores (in various proportions), which has non-negligible consequences and may be unrealistic. This is why we also consider distributions which are equal to zero out of a given interval (support of the distribution strictly included in [0.7,1.3]. In this case, for a given pore with average heterogeneity $h_{pore}$, $h$ is distributed uniformly (simplest model) over the interval [$h_{pore}$-$\delta_{domain}$; $h_{pore}$ + $\delta_{domain}$], the average value $h_{pore}$ being chosen uniformly over the interval [1-$\delta_{pore}$; 1+ $\delta_{pore}$] (see Fig. 1). With such a distribution, all heterogeneity values are not found in all pores. The variance of this distribution is $\delta_{pore}^2/3$.

The simulations are performed in the grand canonical ensemble. The chemical potential of argon is imposed and related to the gas pressure in the reservoir using the ideal gas relation since it is a good approximation at the temperature at which the simulations are performed ($T^*=Tk/\varepsilon_{Ar-Ar}=0.6$, which corresponds approximately to 70K). The saturating chemical potential for the pure fluid at this temperature has been calculated by GEMC:[35] $\mu^{sat}$ = -7.92 $\varepsilon_{Ar-Ar}$. All adsorption/desorption results are given as a function of the reduced chemical potential $\mu^* = \mu/\varepsilon_{Ar-Ar}$. The relative pressure $P/P^{sat}$ can be easily deduced from $\mu$-$\mu^{sat}$. The correspondence is shown in the first figure only (Fig. 2, bottom and top axis). Long simulations are performed in order to reduce the uncertainties down to 7%, taking into account the correlation between successive configurations ($10^8$ Monte Carlo trials per argon where required).[34]

The last point to be discussed is the choice of the number of domains and pores in our calculations. Figure 2 shows the adsorption/desorption isotherms obtained by the multiscale approach (to be described in the next section) for a single pore and various numbers of domains in this pore. As can be seen, for 500 or 1000 domains the curves are almost superimposed, whereas for 100 domains some differences appear. Furthermore, for 100 domains, the adsorption/desorption curve is sensitive to the special choice made for the chain of heterogeneities (not shown), which is not the case for 1000 domains per pore. This last value seems to produce results close to the thermodynamic limit. Note that it corresponds to pore approximately 7 μm long. For the same reason, we take 1000 pores in our system. The total number of domains in our model is then $10^6$.

**MULTISCALE APPROACH**

The fluid adsorption properties in the porous material are calculated by the Grand Canonical Monte Carlo (GCMC) method.[36,37] Of course, the brute force simulation in a pore of few micrometers is out of computer capabilities. On the other hand, performing the calculations in each domain independently of the others is a crude approximation. It is possible to perform a parallel calculation where each processor treats a single domain, taking into account the fluid configuration in the rest of the pore (information given by the other processors). The number of required processor is however too large. Reasonable approximations are possible. For instance, in most cases, the fluid-fluid interaction is short range. If the domains are large enough, it can be assumed that the fluid properties in a given domain depend essentially on the fluid state in the neighboring domains (two in the simplest case of tubular pores). It is possible to be even more restrictive, since it is actually sufficient to know the fluid state within a distance of the order of the range of the interaction (for instance $2.5\sigma$ in our case). These small domains in the immediate vicinity of the central domain under study play the role of boundaries, and are intended to reproduce the influence of the rest of the pore. The second approximation is to consider that if the fluid is far from its critical point, the correlations are short range, and then the relevant property of the fluid in the boundary domains is its density distribution and fluctuations, which depends only on the chemical potential imposed to the system. As a consequence, each domain may be treated independently. The difference with the independent domain theory is however the presence of these boundaries and their effect on fluid adsorption in the domain. Considering the essentially cylindrical shape of the pores, and depending on the value of the chemical potential $\mu$, the fluid in the neighboring domains may be in one or two states. For low $\mu$, the fluid is necessarily in the so-called gaslike state (fluid adsorbed at the walls), denoted G. For high $\mu$ the fluid is necessarily in the liquidlike state (fluid filling the pore), denoted L. For intermediate values of $\mu$ the fluid may be in either the gaslike or the liquidlike states. As a consequence, the fluid properties in each domain depend on temperature $T$, $\mu$, and the fluid state in the two boundaries: G-G, G-L, L-G, and L-L. For further technical details, see Ref [34].

The adsorption/desorption curves have now to be determined for any value of the heterogeneity parameter $h$. They are obtained by linear interpolation of the results obtained for $h$ = 0.7, 0.8, 0.9, 1.0, 1.1, 1.2 and 1.3 in a previous work.[34] Depending on the boundary conditions, the curves are reversible (G-L boundaries) or present a hysteresis (G-G and L-L boundaries). In the case of G-L boundaries, the system always presents a meniscus, which moves forward during adsorption and recedes during desorption (at coexistence). For L-L boundary conditions, two menisci fill the pore during adsorption until they merge (at coexistence). Upon desorption, a bubble has to nucleate somewhere in the pore. This delays the desorption (nucleation barrier), which takes place below coexistence. For G-G boundaries, it is the adsorption which has to proceed via nucleation of a meniscus (above the coexistence), while upon desorption (which occurs at coexistence) two menisci are present due to the gaslike boundaries, which simply recede until they merge. Calculations show that the boundaries do not affect the main adsorption and desorption branches, but only the path followed by the system (positions of the transitions). In other words, the boundaries affect the nucleation barriers, not the grand potential of the fluid. When increasing the heterogeneity parameter $h$, hysteresis narrows and shifts to lower $\mu$. The vertical jumps between gaslike and liquidlike branches are considered as discontinuities. The adsorption properties are then described by two continuous surfaces in the ($\mu$, $h$) plane, and several lines of discontinuities associated to vertical adsorption or desorption. Interpolation between the various ($\mu$, $h$) points calculated by simulation is made in the simplest way (linear) for each of the two surfaces.

The adsorption/desorption isotherms for the whole system are calculated as follows. For low $\mu$, the adsorption in each domain is single valued, and does not depend on the fluid state in the rest of the pore. The amount adsorbed is simply the sum over all domains. Note that in this case, all domains are in the gaslike state and follow the isotherm with G-G boundaries. The chemical potential is then increased by small amounts until one domain reaches its limit of stability. This domain jumps to the liquidlike branch. This has consequences for the two neighboring domains, which boundaries are now G-L and L-G. Their state has then to be reconsidered according to their dissymmetric G-L isotherms. They may stay in the gaslike state (adsorption limited to the initial domain), or switch to the liquidlike state (propagation of the meniscus created in the initial domain). In this case, the next domains need now to be reconsidered, and so on until the meniscus stops (end of the avalanche). The amount adsorbed is given by the sum over the domains. The chemical potential in then increased again, and the same procedure is applied until saturation is reached. The desorption and scanning isotherms are obtained in a similar way.

**SIMULATION RESULTS AND DISCUSSION**

The adsorption, desorption and first scanning descending and ascending isotherms are calculated for various values of the heterogeneity distribution parameters $\sigma_{pore}$, $\sigma_{domain}$, $\delta_{pore}$, and $\delta_{domain}$: 0, 0.05, 0.1, and 0.2. The curves are presented in arrays for clarity (Figs 3-10). For each hysteresis, 3 scanning descending and 3 scanning ascending curves are shown (when possible) in the lower, intermediate, and upper parts of the loop. They are given in separate figures. In all cases, the amount adsorbed is divided by the number of domains ($10^6$), and given as a function of the reduced chemical potential $\mu/\varepsilon_{Ar-Ar}$.

**Gaussian distribution for pores open at both ends**

The simplest case concerns a set of identical, perfectly homogeneous pores ($\sigma_{pore} = \sigma_{domain} = 0$) open at both ends (lower-left panel in Fig. 3). The main hysteresis is identical to the one obtained in a single domain for $h = 1$. The vertical branches are unphysical because, in the grand canonical ensemble, the system jumps from the gaslike to the liquidlike state without intermediate densities. As a consequence, the exploration of the hysteresis loop in this grand canonical situation is impossible.

The introduction of a small disorder between pores ($\sigma_{pore} \neq 0$) has important implications. The case $\sigma_{pore} = 0.05$ and $\sigma_{domain} = 0$, shown in Fig. 3, corresponds to a Gaussian distribution of independent, homogeneous pores with various fluid/wall interactions. The adsorption and desorption branches are now continuous and smooth. They remain essentially parallel. Three descending (Fig. 3) and ascending (Fig. 4) scanning curves are shown. They are essentially straight and parallel to the main gaslike and liquidlike branches. Starting from the main adsorption or desorption branches at a given level of filling of the system, they reach the opposite branch straightforwardly (so-called "crossing" ascending and descending scanning curves). As expected, these properties are those given by the independent domain theory.

The introduction of a small width $\sigma_{domain}$ has different consequences, as shown in Figs 3 and 4 for the case $\sigma_{pore} = 0$ and $\sigma_{domain} = 0.05$, corresponding to a collection of statistically identical heterogeneous pores. The main adsorption and desorption branches are continuous and almost parallel, as previously, but they are shifted to lower $\mu$, and the slopes are significantly steeper. This difference is a direct consequence of the interconnection between the various domains which allows the appearance and propagation of menisci (avalanches). The main differences appear in the scanning curves. The initial portions of the ascending and descending scanning curves are essentially parallel to the main gaslike and liquidlike branches, but they deviate at the end. The ascending curves get steeper before they reach the opposite adsorption branch (crossing ascending scanning curves). The effect is stronger for the descending curves which drop parallel to the desorption branch, and reach the lower closure point of the hysteresis (descending converging scanning curves). This is interpreted as follows: upon desorption (scanning descending curves), the less favorable domains which have not been filled during adsorption (gaslike bubbles in the material) significantly favor the complete desorption of the system compared to the main desorption loop (from the saturated system). This is again a consequence of the interconnection between domains. It is emphasized that this difference of behavior between ascending and descending scanning curves (crossing versus converging) is quite remarkable. Despite it has been observed experimentally for $N_2$ adsorption at 77K in SBA-15 samples,[38] few models are able to reproduce it. For instance, the independent domain theory, networks approaches, and mean-field density functional theories applied to lattice gases, generally predict the same behavior (crossing or convergence) for both ascending and descending curves.[31,32,38-42] Our approach then allows to reproduce a wider variety of behaviors.

Another interesting case is the combined distribution for pores and domains with equal variances ($\sigma_{pore} = \sigma_{domain} = 0.05$, Figs. 3-4). Here again the main adsorption and desorption branches are continuous, but contrarily to the previous cases, the branches are not parallel anymore: the adsorption is smoother than desorption. The scanning curves exhibit an intermediate behavior between the previous cases: they are not parallel to the gaslike and liquidlike branches, and reach the opposite branch. Note that the introduction of some disorder between pores makes the descending curves more and more crossing-like.

Increasing $\sigma_{pore}$ and $\sigma_{domain}$ enhance or eventually completely changes the previous observations (see Figs. 5-6). For instance, for perfectly cylindrical pores ($\sigma_{domain} = 0$), an increase in $\sigma_{pore}$ induces an enlargement of the hysteresis, while the branches remain parallel. The ascending and descending scanning curves remain parallel to the gaslike and liquidlike branches (essentially horizontal), and reach the opposite adsorption or desorption curves. For statistically identical heterogeneous pores ($\sigma_{pore} = 0$), increasing the value of $\sigma_{domain}$ makes the hysteresis more and more triangular, with steeper desorption and smoother adsorption. Note the kink in adsorption. It corresponds to the limit of stability of the gaslike branch for the most attractive domains, which induces nucleation and rapid propagation of meniscus in pores. The descending scanning curves are not parallel to the liquidlike branch, and reach the main desorption branch at the lower closure point. The ascending scanning curves are initially parallel to the gaslike branch, but finally increase rapidly to meet the adsorption branch at the upper closure point of the main hysteresis loop. When both distributions are combined ($\sigma_{pore} \neq 0$ and $\sigma_{domain} \neq 0$), increasing $\sigma_{pore}$ makes the adsorption and desorption branches smoother. On the other hand, when increasing $\sigma_{domain}$ the hysteresis loop tends to converge toward a triangular shape which is almost independent on $\sigma_{pore}$. In other words, the disorder within the pores has stronger effect than that between the pores.[8] Inspection of hysteresis shows that, for $\sigma_{domain}$ = 0 and 0.05, the hysteresis exhibit parallel branches with essentially crossing scanning curves, while for $\sigma_{domain}$ = 0.1 and 0.2, the hysteresis are essentially triangular, with converging scanning curves. A qualitative change in hysteresis structure then takes place for $\sigma_{domain}$ between 0.05 and 0.1.

**Uniform distribution for pores open at both ends**

Figures 7-8 show the adsorption/desorption isotherms and scanning curves for the uniform distributions characterized by $\delta_{domain}$ and $\delta_{pore}$. As for Gaussian distributions, the introduction of differences between homogeneous pores ($\delta_{domain}$ = 0, $\delta_{pore} \neq 0$) makes the adsorption and desorption curves smoother while they remain almost parallel. The scanning adsorption and desorption are parallel to the gaslike and liquidlike branches, and meet the opposite adsorption or desorption branch. On the other hand, for a distribution of statistically identical heterogeneous pores ($\delta_{pore}$ = 0, $\delta_{domain} \neq 0$), the introduction of a small distribution of heterogeneities ($\delta_{domain}$ = 0.1) has almost no effect on the main hysteresis loop, while for stronger disorder ($\delta_{domain}$ = 0.2), the adsorption branch becomes smoother while the desorption branch remains vertical. As can be seen, each time the adsorption or desorption branch is vertical, the desorption or adsorption scanning curves do not exist. For $\delta_{pore}$ = 0 and $\delta_{domain}$ = 0.2, the scanning desorption curves are initially almost parallel to the liquidlike branch, but finally decrease rapidly to meet the main desorption branch at the lower closure point of hysteresis.

The introduction of both kinds of heterogeneities exhibit interesting behaviors. For moderate heterogeneities along the pores ($\delta_{domain} = 0.1$), the pore distribution essentially produces smooth, almost parallel adsorption/desorption branches, with scanning curves parallel to the corresponding gaslike and liquidlike branches, and crossing the hysteresis loop. In this case, the pore distribution dominates, and the system behaves essentially accordingly to the independent pore model. Interestingly, for a given pore distribution amplitude ($\delta_{pore}$), the presence of small heterogeneities along each pore ($\delta_{domain} = 0.1$) makes the adsorption and desorption branches steeper compared to the case $\delta_{domain} = 0$, and slightly displaced towards lower $\mu$ due to easier nucleation on the most attractive sites during adsorption, and delayed meniscus recession due to the same attractive sites upon desorption. On the other hand, for larger heterogeneities along the pores ($\delta_{domain} = 0.2$), the desorption branches are steeper, while the adsorption branches are smoother than for $\delta_{domain} = 0.1$. The scanning curves also exhibit drastic change for $\delta_{domain} = 0.2$, since they have a tendency to reach the upper and lower hysteresis closure point.

Inspection of hysteresis shows that, for $\delta_{domain} = 0$ and 0.1, the hysteresis exhibit essentially parallel branches with crossing scanning curves, while for $\delta_{domain} = 0.2$, the hysteresis are triangular, with essentially converging scanning curves. A qualitative change in hysteresis structure then takes place for $\delta_{domain}$ between 0.1 and 0.2. Such a behavior was previously noticed for Gaussian distributions, for $\sigma_{domain}$ between 0.05 and 0.1. Since the variance of the flat distribution is $\delta_{domain}{}^2/3$, the qualitative change in hysteresis occurs on the same variance interval of pore heterogeneity for both distributions.

**Comparison between pores open at one or at both ends**

The case of pores closed at one end is now considered. All results obtained for Gaussian distributions are identical to the case where pores are open at both ends, and are thus not shown. On the other hand, in the case of uniform distributions, the isotherms exhibit differences when pores are closed at one end (see Fig. 9-10). The differences appear only for adsorption, which occurs more rapidly for pores closed at one end (easier nucleation of the meniscus). After further increase in chemical potential beyond the point where a meniscus can nucleate in the most attractive domains of the pores, the two adsorption curves merge. The most striking difference is for $\delta_{domain} = 0$, in which case the isotherms become reversible for pores closed at one end. On the other hand, for wide distributions of heterogeneities along the pores ($\delta_{domain} = 0.2$) the effect of pore ends is negligible. In other words, the larger the support of the distribution ($2\delta_{domain}$), the smaller the influence of pore ends.

Why do we observe a difference between closed and open pores for uniform truncated distributions and not for Gaussian distributions? The difference between pores open at one or at both ends is that, in the first case, the closed end immediately provides a fluid meniscus. However, to produce a difference in adsorption, this meniscus has to be able to propagate at least in some of the pores. This propagation occurs only in pores for which all domains have a fluid/wall heterogeneity parameter higher than a threshold value imposed by $\mu$. This is possible in the case of the uniform and truncated distributions, which explains that a difference is observed between open and closed pores. On the other hand, in the case of the Gaussian distribution, all pores contain several domains of the lowest fluid/wall parameter (even though the probability might be small), which always impedes the propagation of the meniscus in the hysteresis region ($\mu < -8.2$). In the thermodynamic limit, the propagation of the meniscus from the closed end is negligible in all pores, and hysteresis loop is identical whatever the pores are closed or open. It is then the extreme values of the disorder in the pores which seems to be relevant to explain the possible difference between pores open at one or at both ends.

**CONCLUSION**

The adsorption of a simple fluid in a mesoporous silica substrate made of independent heterogeneous tubular pores (as MCM-41 or porous silicon) is addressed by a multiscale approach based on GCMC molecular simulation. This framework allowed calculation of adsorption/desorption hysteresis loop and scanning curves for large scale porous material models (1000 pores, several µm long), inaccessible to conventional molecular simulations. The natural heterogeneity in the porous material is characterized by a single parameter, essentially related to the fluid/wall interaction intensity. This disorder is supposed to follow a double distribution, which characterizes the heterogeneity between pores and within each pore. Two types of distributions are used: Gaussian or uniform and truncated.

The results first show the importance of a large number of various domains (heterogeneities) to produce realistic hysteresis loops presenting a multiplicity of metastable states. The models obtained with extremely sharp distributions produce vertical hysteresis, where scanning curves cannot be drawn. Furthermore, the interconnection between the domains, despite simply chainlike in our case, has proven to be of greatest importance to determine the hysteresis loop shape. This can easily be seen by exchanging the values of $\sigma_{pore}$ and $\sigma_{domain}$ or $\delta_{pore}$ and $\delta_{domain}$. For instance, $\sigma_{pore} \neq 0$ and $\sigma_{domain} = 0$ corresponds to a collection of independent domains, while $\sigma_{pore} = 0$ and $\sigma_{domain} \neq 0$ corresponds to interconnected domains (heterogeneous tubular pore). As a general trend, for a small disorder within the pores, the hysteresis essentially presents parallel adsorption/desorption branches, whatever the level of disorder between pores. The scanning curves are essentially parallel to the gaslike and liquidlike branches (crossing scanning curves). This behavior is reminiscent of MCM-41 adsorption experiments. On the other hand, for stronger disorder within pores, the hysteresis is essentially triangular, with desorption steeper than adsorption, and scanning curves converging to lower or higher closure points of the hysteresis. This behavior is reminiscent of adsorption in porous silicon. This qualitative change in behavior occurs for a disorder variance within pores around 0.3-1.0 percent, whatever the distribution shape. It also shows that the disorder within the pores is more important in determining hysteresis shape than the disorder between pores. This work shows that a tree-dimensional network of interconnections is not required to produce a triangular hysteresis, the main origin being the chemical heterogeneity in the pores. It would however be very interesting to study the effect of possible interconnections or any other "interaction mechanism" [33] between pores to compare with the effect of disorder. Future work will also be devoted to the effect of a true pore size distribution to improve comparison with experiments. It was also shown that, contrary to most of usual calculations, our approach is able to produce hysteresis with ascending and descending scanning curves with mixed behavior, i.e. crossing and converging, which was observed in some experimental systems.

The last important conclusion is that comparison between pores open at one or at both ends only exhibit differences if the heterogeneity distribution in the pores is zero out of an interval which is not too wide. In other words, the disorder distribution in the pores should not have long tails. In this case, the consequence of having one of the pore ends closed is that capillary condensation is slightly favored in the first steps, but finally reaches the adsorption branch for pores open at both ends. The hysteresis is then triangular. For extremely homogeneous pores, the hysteresis is significantly reduced, and eventually suppressed for perfectly homogeneous pores. This observation suggests that if MCM-41 pores are quite homogeneous, then they are most probably open at both ends. Another important conclusion from this work in that no difference is observed upon desorption, whatever the pores are open or closed. How does this compare to experiments, performed essentially in porous silicon?[7,33] The authors observe differences between open/closed pores upon adsorption and desorption, which could not be reproduced by our approach. This might be an argument in favor of the effect of the porous matrix elasticity suggested by the authors[33] (which is not taken into account in our study).

This work being a systematic study of the influence of disorder in mesoporous silica made of independent heterogeneous tubular pores, it is expected to be useful to experimentalists interested in the effect of the level of disorder inside and between the pores to interpret experimental adsorption/desorption experiments in MCM-41, porous silicon, or any material made of non-connected pores. It is however emphasized that this study focused on a special case of disorder (modulation of fluid/wall interaction), and therefore conclusions might be different in the presence of other sources of disorder, like pore size or structural variations. Such effects will be studied in future work.

ACKNOWLEDGMENT It is a pleasure to acknowledge A. Delville, E. Kierlik, P. Porion and G. Tarjus for fruitful discussions. The Institut de Développement des Ressources en Informatique Scientifique (IDRIS-CNRS, Orsay, France) is gratefully acknowledged for providing computer capabilities.

FIGURE CAPTIONS

FIG.1. Schematic representation of the mesoporous substrate made of non-interconnected heterogeneous pores. The amplitude of the heterogeneity is materialized by the thickness of the domains. The heterogeneity is described by two distributions which characterize the heterogeneity within the pores ($\sigma_{\text{domain}}$ or $\delta_{\text{domain}}$) and between the pores ($\sigma_{\text{pore}}$ or $\delta_{\text{pore}}$). Two distributions are considered: the Gaussian (a) and uniform truncated (b).

FIG.2. Adsorption/desorption isotherms obtained in the multiscale approach at $T^* = 0.60$ for argon confined in a single heterogeneous tubular pore (Gaussian distribution with $\sigma_{\text{domain}} = 0.2$ and $h_{\text{pore}} = 1.0$, see text). The number of domains is given in the figure. The amount adsorbed is normalized to the number of domains, and given as a function of the reduced chemical potential $\mu/\varepsilon_{\text{Ar-Ar}}$ (bottom axis) or relative pressure $P/P^{\text{sat}}$ (top axis).

FIG.3. Main adsorption/desorption hysteresis loop and first descending scanning curves obtained in the multiscale approach at $T^* = 0.60$ for argon confined in a mesoporous silica sample (solid lines). The doted lines are guides to the eye corresponding to vertical jumps. The heterogeneity distribution is Gaussian, with parameters $\sigma_{\text{pore}}$ and $\sigma_{\text{domain}}$ given in the figure (see text and Fig. 1 for definition). From left to right: increasing $\sigma_{\text{domain}}$, characterizing disorder within the pores; from bottom to top: increasing $\sigma_{\text{pore}}$, characterizing the disorder between pores.

FIG.4. Same as Fig. 3 with the first ascending scanning curves.

FIG.5. Same as Fig. 3 for stronger disorder (larger values of $\sigma_{\text{pore}}$ and $\sigma_{\text{domain}}$).

FIG.6. Same as Fig. 5 with the first ascending scanning curves.

FIG.7. Main adsorption/desorption hysteresis loop and first descending scanning curves obtained in the multiscale approach at $T^* = 0.60$ for argon confined in a mesoporous silica made of pores open at both ends. The doted lines are guides to the eye corresponding to vertical jumps. The heterogeneity distribution is uniform truncated, with parameters $\delta_{\text{pore}}$ and $\delta_{\text{domain}}$ given in the figure (see text and Fig. 1 for definition). From left to right: increasing $\delta_{\text{domain}}$, characterizing disorder within the pores; from bottom to top: increasing $\delta_{\text{pore}}$, characterizing the disorder between pores.

FIG.8. Same as Fig. 7 with the first ascending scanning curves.

FIG.9. Same as Fig. 7 for pores closed at one end. For comparison, the main hysteresis loop for pores open at both ends is given as doted lines.

FIG.10. Same as Fig. 9 with the first ascending scanning curves.

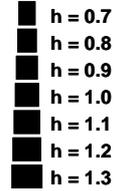
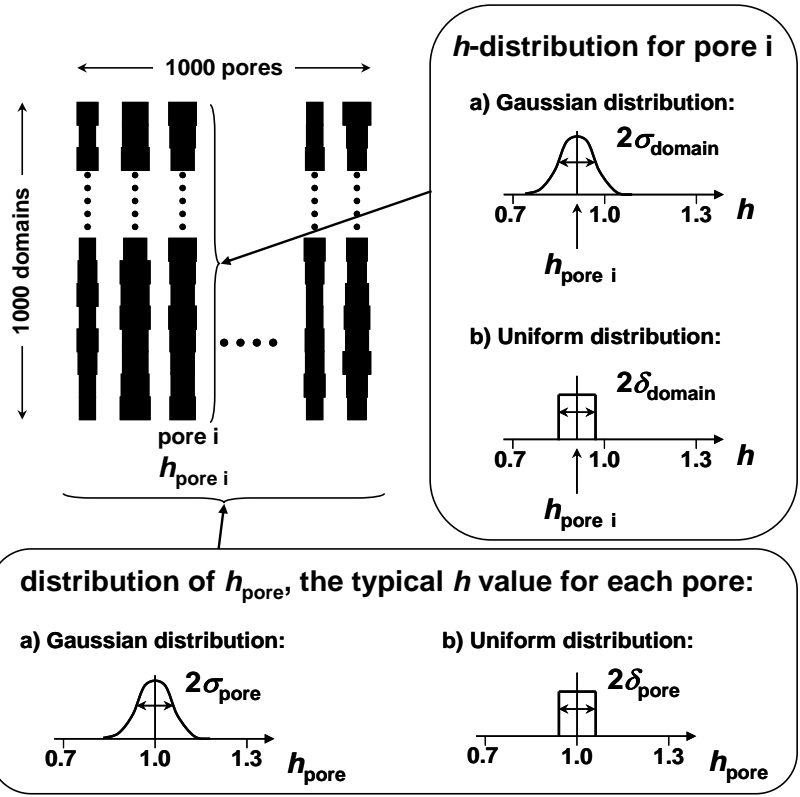

Figure 2

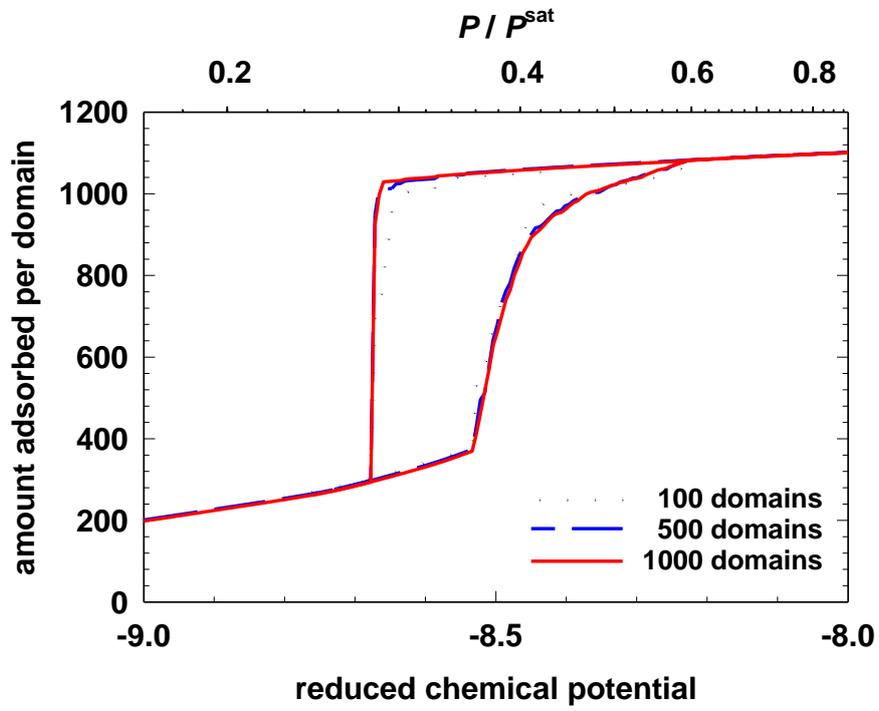

Figure 3

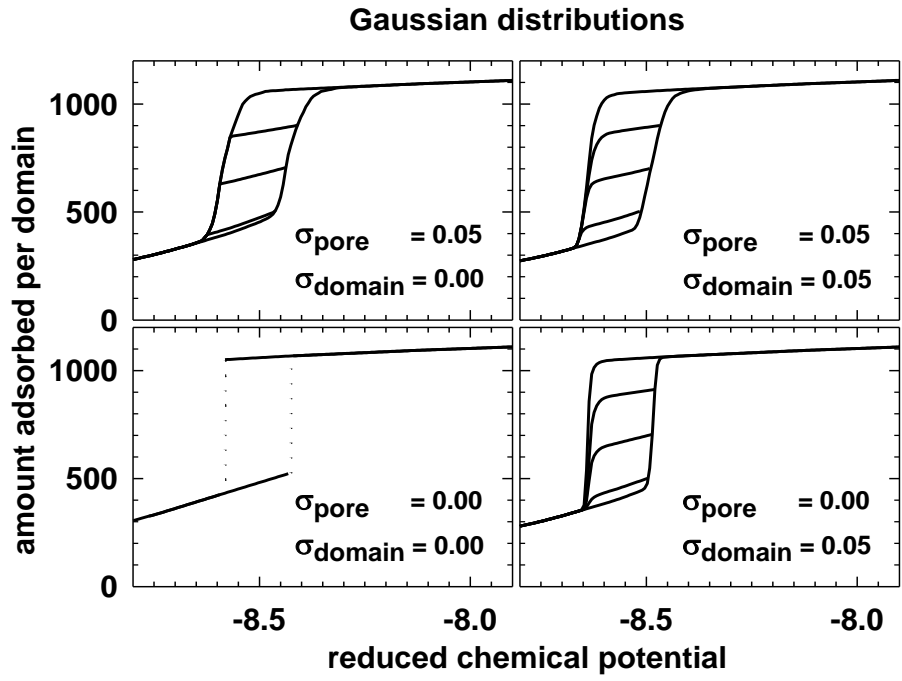

Figure 4

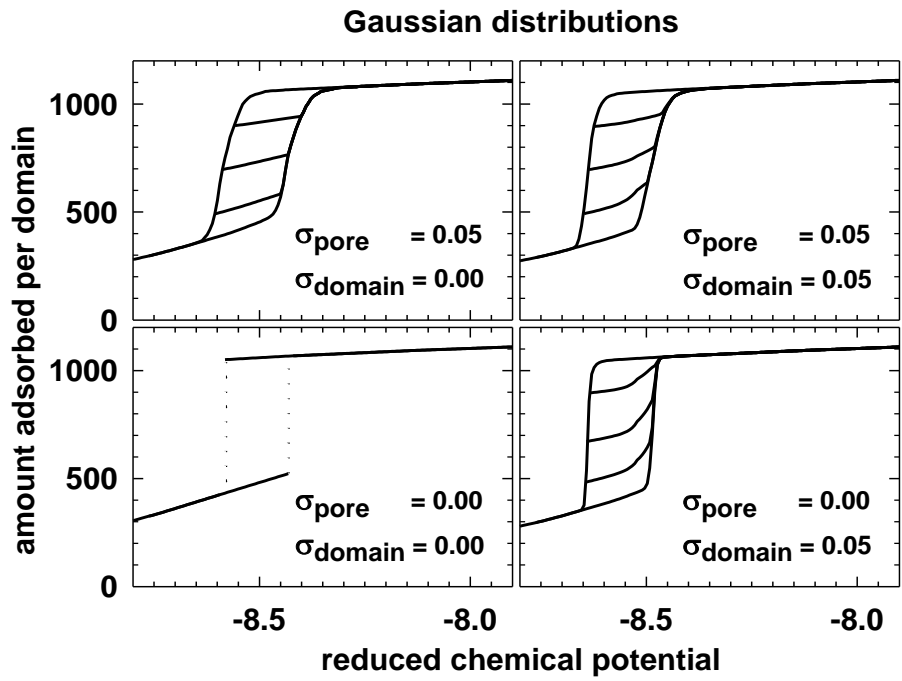

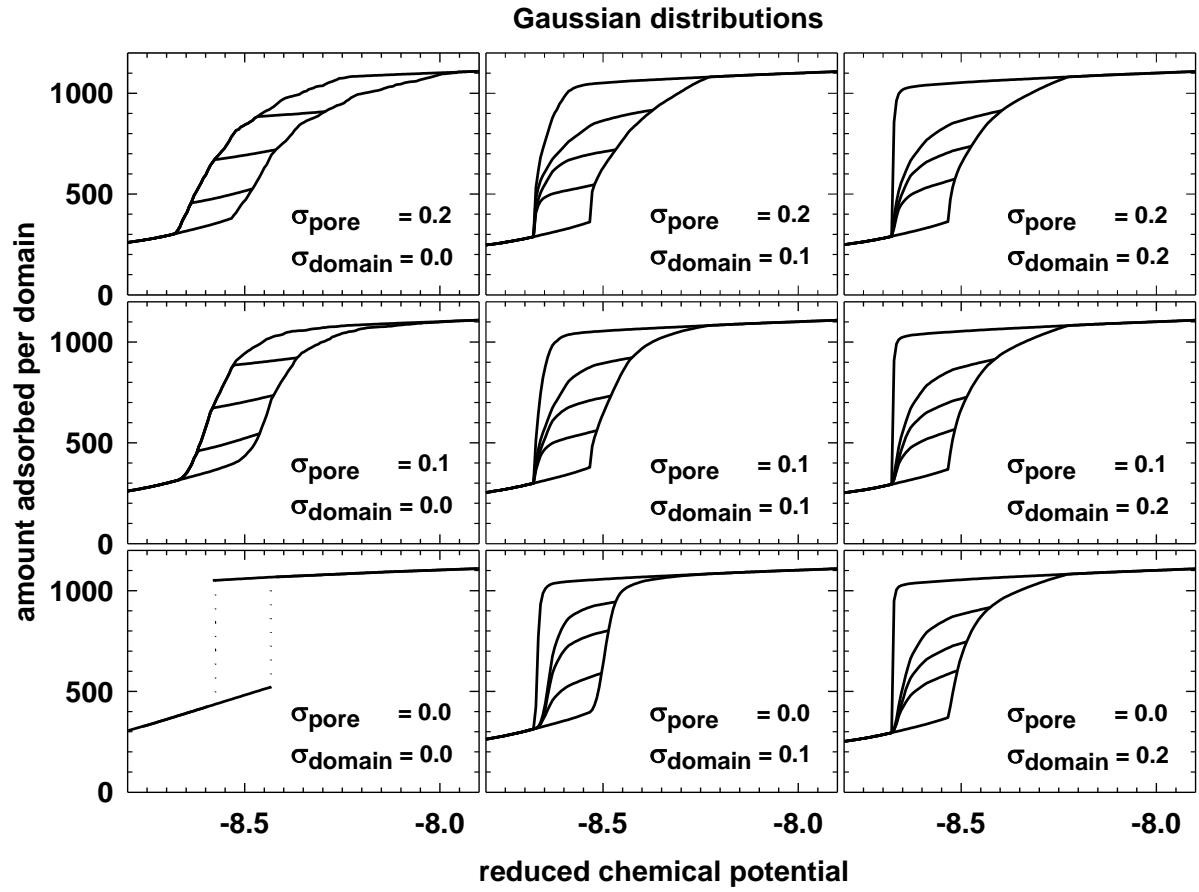

Figure 6

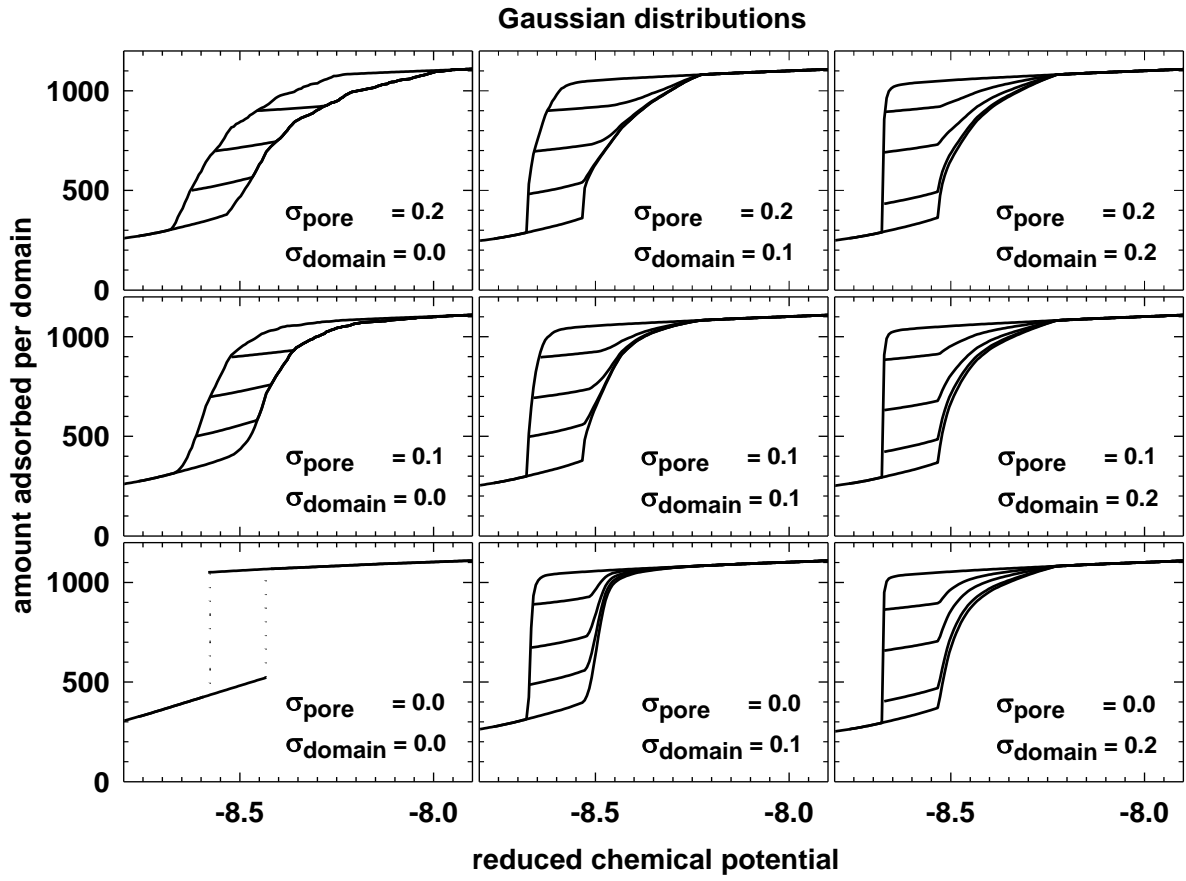

Figure 7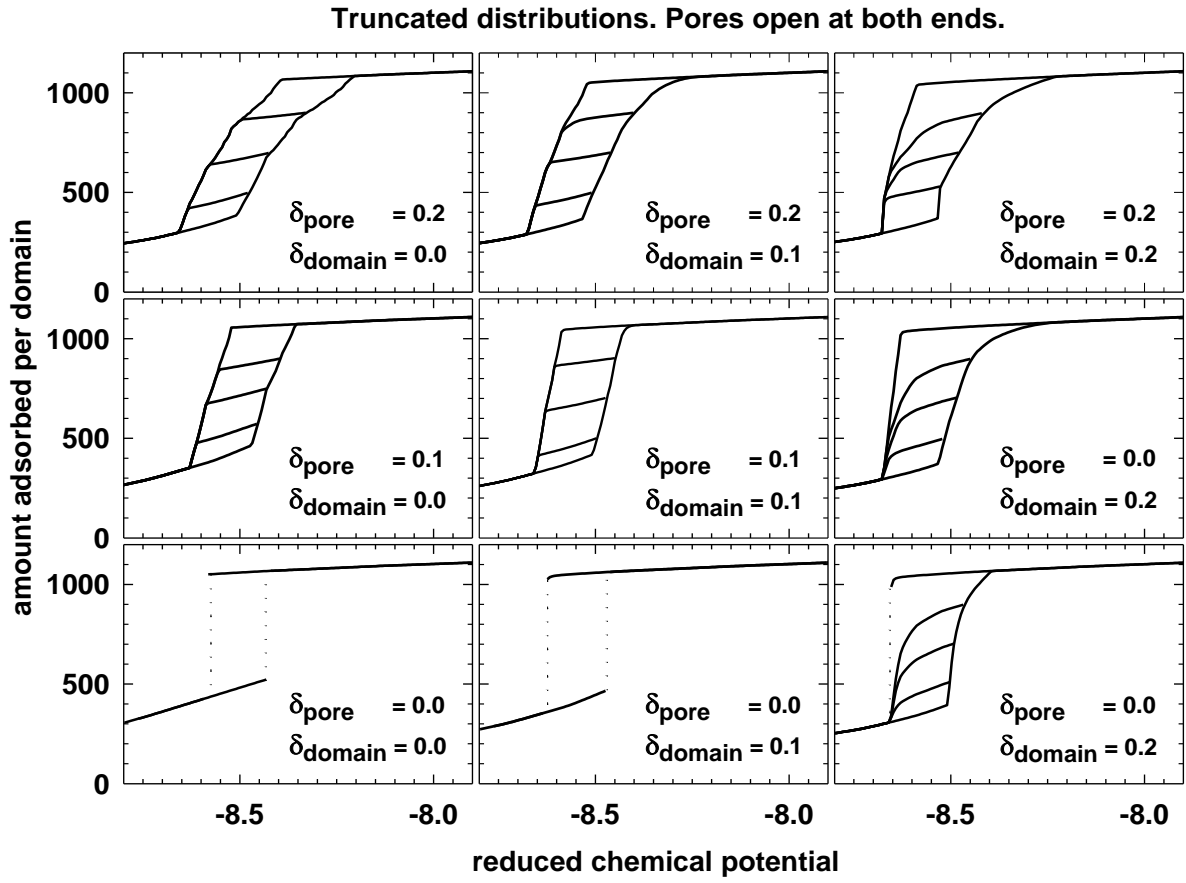

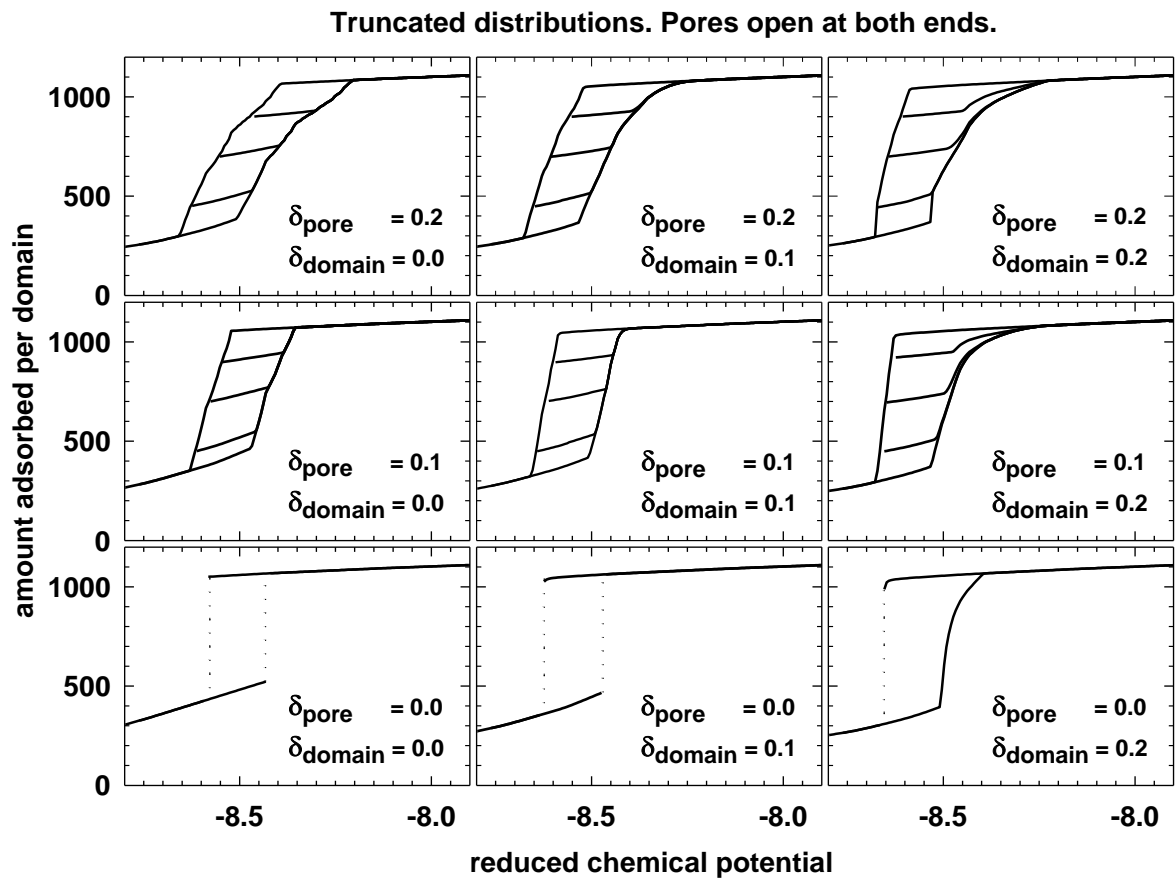

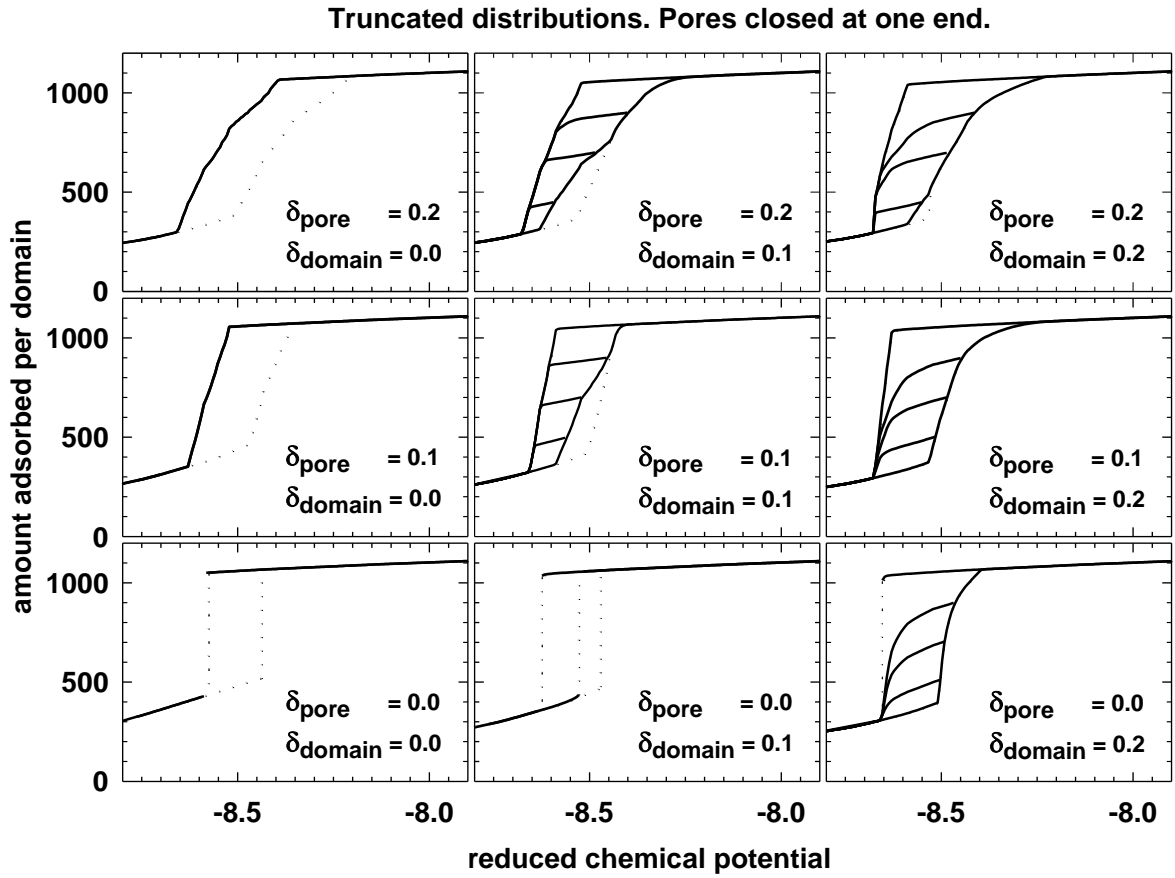

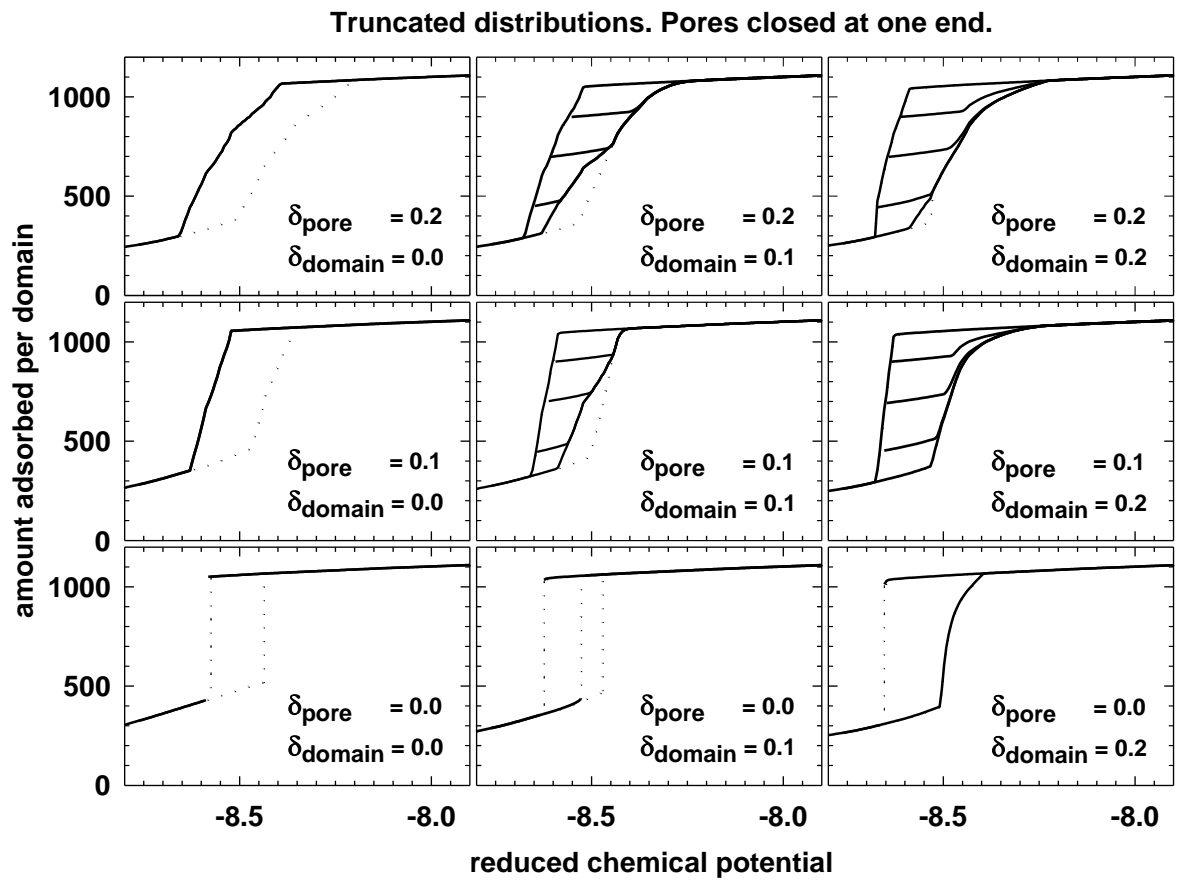

SYNOPSIS TOC (Word Style "SN_Synopsis_TOC").

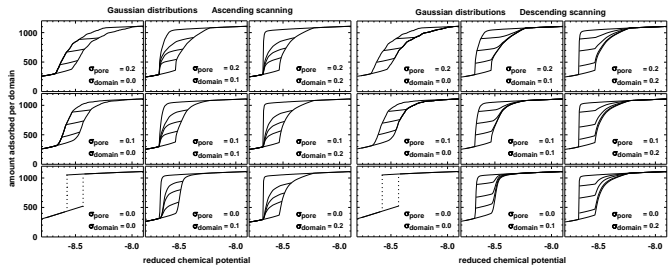